\documentclass[prl,aps,superscriptaddress,twocolumn,floatfix]{revtex4}
\usepackage{latexsym}
\usepackage{amsfonts}
\usepackage{amssymb}
\usepackage{amsmath}
\usepackage{graphicx}
\begin{document}
\title{Spectroscopic Evidence for Anisotropic S-Wave Pairing
Symmetry in $\bf MgB_2$}
\date{\today}
\author{P. Seneor}
\author{C.-T. Chen}
\author{N.-C. Yeh}

\affiliation{Department of Physics, California Institute of
Technology, Pasadena, CA 91125}
\author{R. P. Vasquez}
\author{L. D. Bell}

\affiliation{Center for Space Microelectronics Technology, Jet
Propulsion Laboratory, California Institute of Technology,
Pasadena, CA 91109}

\author{C. U. Jung}
\author{Min-Seok Park}
\author{Heon-Jung Kim}
\author{W. N. Kang}
\author{Sung-Ik Lee}

\affiliation{National Creative Research Initiative Center for
Superconductivity and Department of Physics, Pohang University of
Science and Technology, Pohang 790-784, Republic of Korea}

\begin{abstract}
Scanning tunneling spectroscopy of superconducting MgB$_2$ ($T_c =
39$ K) were studied on high-density pellets and c-axis oriented
films. The sample surfaces were chemically etched to remove
surface carbonates and hydroxides, and the data were compared with
calculated spectra for all symmetry-allowed pairing channels. The
pairing potential ($\Delta _k$) is best described by an
anisotropic $s$-wave pairing model, with $\Delta _k = \Delta _{xy}
\sin ^2 \theta _k + \Delta _z \cos ^2 \theta _k$, where $\theta
_k$ is the angle relative to the crystalline c-axis, $\Delta _z
\sim 8.0$ meV, and $\Delta _{xy} \sim 5.0$ meV.
\end{abstract}
 \maketitle

Since the discovery of superconductivity in MgB$_2$ at a
superconducting transition temperature $T_c \sim 39$
K~\cite{ref1}, a number of reports
\cite{ref1,ref2,ref3,ref4,ref5,ref6,ref7} have suggested that this
hole-doped layered superconductor \cite{ref8,ref9,ref10} may be
consistent with conventional BCS $s$-wave pairing. On the other
hand, muon spin rotation ($\mu$SR) studies of MgB$_2$ have found
that the temperature dependence of the magnetic penetration depth
is suggestive of unconventional pairing symmetry with nodes in the
superconducting order parameter \cite{ref11}. To address the issue
of the pairing symmetry in this new superconductor, possible
complications by disorder or surface impurities must be
considered. Indeed, recent x-ray photoemission spectroscopy (XPS)
studies have revealed that MgCO$_3$ and Mg(OH)$_2$ exist on the
surface of as-grown MgB$_2$ \cite{ref12}. It is therefore
important to understand how these surface impurity phases may
contribute to surface-sensitive experiments such as the scanning
tunneling spectroscopy (STS) \cite{ref4,ref5,ref6} and
point-contact measurements \cite{ref7} of the quasiparticle
spectra. In particular, existing STS data on as-grown
polycrystalline MgB$_2$ \cite{ref5} exhibited ``V-shape''
differential conductance ($dI_{NS}/dV$) versus voltage ($V$) plots
near zero-bias (i.e. the Fermi level $E_F$), with rounded
``humps'' rather than sharp peaks at the gap values ($V = \pm
\Delta /e$) and large residual density of states (DOS) at $E_F$.
Those spectra were fitted with an $s$-wave pairing potential
$\Delta$ broadened by disorder parameterized as $\Gamma$, and a
large ratio of $(\Gamma / \Delta) \sim 60\%$ was suggested
\cite{ref5}. For comparison, in cuprate superconductors the
V-shape conductance spectra near $E_F$ for quasiparticle tunneling
along the c-axis are known to be the signature of the
$d_{x^2-y^2}$ pairing symmetry
\cite{ref13,ref13b,ref14,ref15,ref16}, and strong directionality
in the quasiparticle spectra has been observed
\cite{ref13,ref13b,ref14}. In particular, a zero-bias conductance
peak (ZBCP) \cite{ref17,ref18} can occur if quasiparticles are
incident close to the $\{110\}$ nodal direction of the
$d_{x^2-y^2}$-wave order parameter. Thus, should the pairing
symmetry be unconventional, the observation of V-shape tunneling
spectra in polycrystalline MgB$_2$ samples associated with certain
grain orientations would be accompanied by frequent occurrence of
ZBCP for other grain orientations. To date, no ZBCP has been found
from vacuum tunneling studies of as-grown MgB$_2$
\cite{ref4,ref5,ref6}. However, a major concern presented by
existing quasiparticle spectra is that the measured gap values
vary widely, and that most values are smaller than that the BCS
prediction \cite{ref4,ref5,ref6,ref7}.

\begin{figure}
  \centering
  \includegraphics[height=5.4in,width=3.173in]{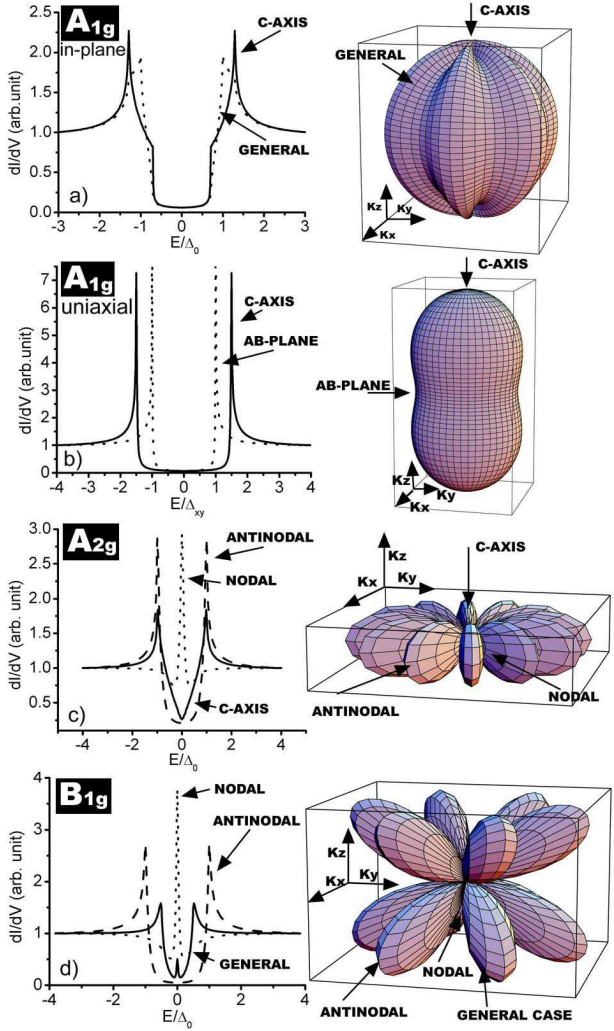}
  \caption{Right panels: Graphical representations for possible order
parameters permitted by the  D$_{6h}$ group symmetry and
spin-singlet pairing. Left panels: Simulated differential
conductance ($G_{NN} dI_{NS}/dV$) vs. voltage ($V$) quasiparticle
tunneling spectra at 4.2 K, assuming $\Delta _0 = 6.5$ meV, for
the following 1D even-parity representations. {\bf (a)} $A_{1g}$,
anisotropic $s$-wave with in-plane anisotropy; {\bf (b)} $A_{1g}$,
anisotropic $s$-wave with uniaxial symmetry; {\bf (c)} $A_{2g}$;
{\bf (d)} $B_{1g}$, or $B_{2g}$ by rotating $B_{1g}$ order
parameter through an angle $(\pi/6)$ relative to $k_z$.
\label{fig1}}
\end{figure}

Our starting point for investigating the pairing symmetry of
MgB$_2$ is to consider all the possible pairing channels based on
group theory. The global symmetry group $\mathcal{G}$ of MgB$_2$
in its normal state can be expressed by $\mathcal{G} = U(1) \times
\mathcal{T} \times SU(2) \times \mathcal{G} _{space}$, where
$U(1)$ is the electromagnetic gauge broken below $T_c$,
$\mathcal{T}$ and $SU(2)$ denote the time-reversal and
spin-rotational symmetries that are generally preserved below
$T_c$ for spin-singlet Cooper pairs, and $\mathcal{G} _{space}$ is
the space group $D_{6h}$ for MgB$_2$. Given that the Cooper pairs
in MgB$_2$ are spin-singlets \cite{ref2} and that no other obvious
symmetry-breaking fields exist below $T_c$ except $U(1)$, the
possible pairing channels can be derived from the even-parity
irreducible representations of $D_{6h}$. For a single-component
superconductor, the relevant pairing channels can be further
reduced to four one-dimensional (1D) even-parity irreducible
representations in $D_{6h}$: $A_{1g}$, $A_{2g}$, $B_{1g}$ and
$B_{2g}$. The pairing potentials $\Delta _k$ for these
representations can be expressed as a function of the momentum
$\vec k$ to the lowest order:
\begin{eqnarray}\label{equ1}
A_{1g}: & \Delta _k = \Delta _0, (\textrm{isotropic \emph{s}})
\nonumber \\
 :& \Delta _k = \Delta _0
\lbrack 1 + \epsilon \cos (6 \phi _k) \rbrack, (\textrm{anisotropic \emph{s}}) \nonumber \\
:&  \Delta _k = \Delta _{xy} \sin ^2 \theta _k + \Delta _z \cos ^2
\theta _k , (\textrm{anisotropic \emph{s}}) \nonumber \\
A_{2g} :& \Delta _k = \Delta _0 \sin ^6 \theta _k \sin (6 \phi
_k), \\ B_{1g} :& \Delta _k =
\Delta _0 \cos \theta _k \sin ^3 \theta _k \sin (3 \phi _k ),\nonumber  \\
B_{2g} :& \Delta _k = \Delta _0 \cos \theta _k \sin ^3 \theta _k
\cos (3 \phi _k ). \nonumber
\end{eqnarray}
Here $\theta _k$ is the angle measured relative to $\hat k _z$,
with $\hat k _z$ parallel to the crystalline c-axis, and $\phi _k$
is measured relative to $\hat k _x$. In addition, $0<\epsilon < 1$
and $\Delta _{xy} \not= \Delta _z$ for the anisotropic $s$-wave
pairing potentials. The graphical representations of these
different pairing potentials are illustrated in
Figs.~\ref{fig1}(a)-(d).

Among different $A_{1g}$-representations, the lowest-order
possibilities include the isotropic $s$-wave order parameter,
anisotropic $s$-wave with 6-fold in-plane modulations, or
anisotropic $s$-wave with uniaxial symmetry, with the latter two
illustrated in Figs.~\ref{fig1}(a) and \ref{fig1}(b). The
lowest-order $A_{2g}$-representation consists of twelve ``lobes''
of alternating phases, and the phases are even under
$k_z$-inversion. For either $B_{1g}$ or $B_{2g}$-representation,
the order parameter consists of twelve lobes with alternating
phases, and the phases are odd under $k_z$-inversion.

To obtain the quasiparticle spectra for all possible pairing
channels with different $\Delta _k$, we consider a crystalline
plane with a normal vector $\hat n _k$ characterized by the
parameters $(\theta _k, \phi _k)$. Defining the direction of an
incident quasiparticle relative to $\hat n _k$ by the parameters
$\hat \ell (\theta _{in}, \phi _{in}) \equiv \hat \ell _{in}$,
which explicitly considers the transverse momentum for the
incident quasiparticles (i.e., a finite ``tunneling cone'')
relative to $\hat n _k$, such that $\theta _{in}$ is primarily
confined between $- \beta$ and $ \beta$, and $0 \le \phi _{in} \le
2 \pi$, we can generalize the theory of Blonder-Tinkham-Klapwijk
(BTK) \cite{ref17,ref18,ref19} to three dimensions (3D), and
compute the tunneling current $I_{NS}$ as a function of the bias
voltage $V$, temperature $T$, tunneling barrier strength $Z$,
tunneling direction $\hat n _k$, and tunneling cone $\beta$:
\begin{eqnarray}\label{equ2}\nonumber I_{NS} & = & G_{NN} \int _0 ^{2 \pi} d \phi _{in} \int _0
^{\pi /2} d \theta _{in} \cos \theta _{in}\ e^{ - \frac{\theta
_{in}^2}{\beta ^2}} \int dE_k \\ & & \times \left[ 1 + A - B
\right] \times \left[ f(E_k - eV) - f(E_k) \right].
\end{eqnarray}
In Eq.~(\ref{equ2}), $G_{NN}$ denotes the normal-state
conductance, $E_k$ is the quasiparticle energy, $A$ and $B$
represent the kernals for Andreev and normal reflection,
respectively, and $f(E_k)$ is the Fermi function
\cite{ref17,ref18}. Thus, the differential conductance spectra
$(dI_{NS}/dV)$-vs.-$V$ can be obtained for given $\hat n _k$ and
$\Delta _k$ using Eqs.~(\ref{equ1}) and (\ref{equ2}). The
representative spectra for high-impedance tunneling barrier $Z =
5$ are shown in the left panels of Fig.~\ref{fig1}(a)-(d).

Except for the $A_{1g}$-representation, the spectral
characteristics for all other representations exhibit strong
directionality (i.e., dependence on the crystalline normal $\hat n
_k$ relative to the average quasiparticle momentum), as manifested
by calculated spectra in the right panels of
Fig.~\ref{fig1}(a)-(d). It is clear that the ZBCP would have been
a common occurrence in the tunneling spectra of MgB$_2$ pellets
had the order parameter been one of the unconventional pairing
channels ($A_{2g}, B_{1g}, B_{2g}$).

To compare the calculated results with experiments, we performed
scanning tunneling spectroscopy on high-density pellets
\cite{ref20,ref21,ref21b} and c-axis textured films of MgB$_2$
\cite{ref22} at 4.2 K. Both the pellets and c-axis films were
fully characterized \cite{ref12,ref20,ref21,ref21b,ref22}, showing
single-phased material with superconducting transition at $T_c =
39.0$ K, sharp magnetization transition widths ($\Delta T_c < 1$ K
for the pellets and $\Delta T_c \sim  0.7$ K for the films), and
nearly 100\% bulk superconducting volume
\cite{ref20,ref21,ref21b,ref22}. According to XPS studies on these
samples \cite{ref12}, the surface  MgCO$_3$ and Mg(OH)$_2$
impurities on the as-grown MgB$_2$ could be mostly removed by
chemical etching, with no discernible etch residues for the
tunneling experiments \cite{ref12}. Tunneling studies were
conducted on the as-grown and etched MgB$_2$ pellets and films at
4.2 K, using a low-temperature scanning tunneling microscope.
Spatially resolved tunneling spectra were taken on over 100
randomly oriented grains of each sample. On each grain, the
spectra were taken under the vacuum tunneling condition and on an
area approximately $(200 nm \times 200 nm)$ in size with
nano-scale surface flatness. A large number of grains were studied
on each sample to ensure sufficient statistical sampling of
different $\hat n _k$ in pellets.

\begin{figure}
  \centering
  \includegraphics[width=3.355in,height=2.34in]{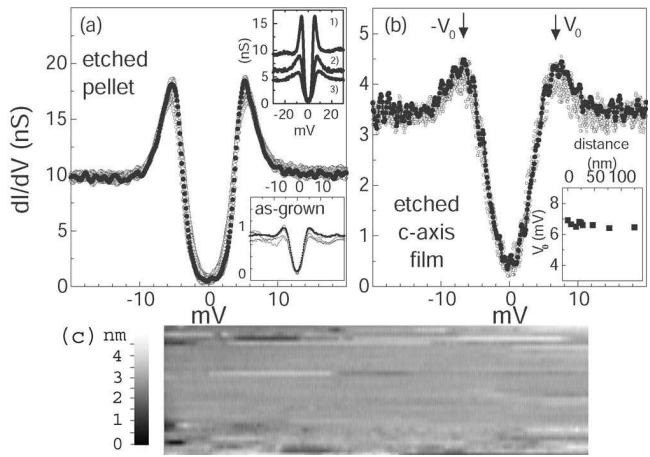}
  \caption{{\bf (a)} Spatially resolved tunneling spectra of a high-density
MgB$_2$ pellet. The main panel and the lower right inset
illustrate data taken at locations $10 \sim 15$ nm apart within
one grain after and before chemical etching, respectively. The
upper left inset shows representative spectra on the etched pellet
with different junction resistance at 20 mV: 1) 108~M$\Omega$, 2)
179~M$\Omega$, 3) 253~M$\Omega$.  The work function for these
spectra is typically $0.1 \sim 1$ eV. {\bf (b)} A series of
tunneling spectra on an etched c-axis film (main panel), showing
long-range spatial homogeneity in the spectral peak-to-peak
energies and a large junction resistance $\sim$~330~M$\Omega$
(inset). {\bf (c)} An image of the surface topography of the
etched sample over an area $(196 nm \times 60 nm)$. The full scale
for the height is 4.7~nm.\label{fig2}}
\end{figure}
Representative tunneling spectra for a MgB$_2$ pellet after
etching are shown in the main panel of Fig.~\ref{fig2}(a), and
those for the same sample before etching are given in the lower
right inset. We note significantly improved spectra after etching,
with long-range spatial homogeneity ($> 400$ nm) within each
grain, which correlated well with the long-range atomic flatness
of the topography as exemplified in Figure 2(c) and also according
to our AFM images, and was in contrast to the strong spatial
variations in both the spectra and topography of MgB$_2$ powder
\cite{ref22b}. Furthermore, the density of states (DOS) nearly
vanished at $E_F$, with a normalized value $\lbrack
(dI_{NS}/dV)_{V=0}/(dI_{NS}/dV)_{V=20 meV} \rbrack \sim 2 \%$.
While the tunneling spectra were homogeneous within each grain
(with lateral dimension $\sim$ a few $\mu$m \cite{ref21}), the gap
values varied from grain to grain in the pellets, ranging from
$\sim 5$ to $\sim 8$ meV. On the other hand, tunneling spectra of
etched c-axis oriented films were homogeneous everywhere. Overall,
no ZBCP was observed among over five hundred spectra taken on all
samples. We therefore conclude that the pairing symmetry must be
of the $A_{1g}$ representation.

\begin{figure}
  \centering
  \includegraphics[height=3in,keepaspectratio=true]{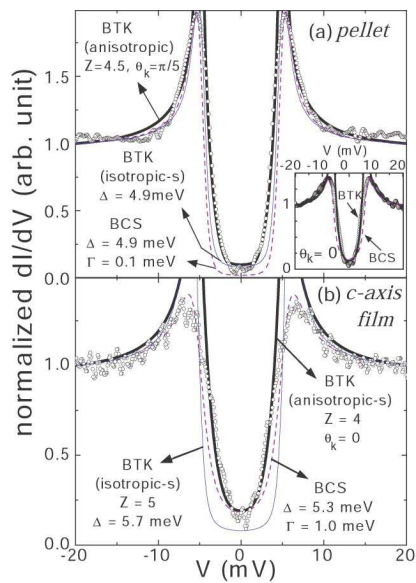}
  \caption{BTK anisotropic and isotropic
$s$-wave fitting, together with the isotropic BCS fitting to
representative spectra of {\bf (a)} an etched MgB$_2$ pellet, and
{\bf (b)} an etched c-axis film. Given empirical values of $\Delta
_{xy}$ and $\Delta _z$, the anisotropic s-wave fitting is only
sensitive to the variation in $\theta _k$ and is insensitive to a
wide range of $\beta$ values that we have tested, from $(\pi/18)$
to $(\pi/2)$. The fitting curves shown have assumed the most
general case with $\beta = \pi/2$.\label{fig3}}
\end{figure}

To identify the correct pairing potential under the $A_{1g}$
representation in Eq.~(\ref{equ1}), we performed the BTK analysis
for both anisotropic and isotropic $s$-wave pairing, as well as
the isotropic BCS fitting to all spectra. The latter involved a
disorder parameter $\Gamma$ for an isotropic gap $\Delta$ with the
density of states $\mathcal{N} (E)$ given by \cite{ref23}:
$$\mathcal{N} (E) = \textrm{Re} \left[ (E - i \Gamma)/\sqrt{(E - i \Gamma)^2
- \Delta ^2 } \right] \propto dI_{NS}/dV.$$ For both BTK and BCS
isotropic $s$-wave fitting, we notice several difficulties. First,
the inclusion of the disorder-induced pair-breaking strength
$\Gamma$ alone cannot fully account for the spectral
characteristics, particularly the line-width and line-shape of the
peaks, as manifested in Figs.~\ref{fig3}(a) and \ref{fig3}(b).
Second, significant variations in the supposedly isotropic pairing
potential must be invoked to account for all data taken on the
pellets. The variation was unlikely the result of bulk
stoichiometric inhomogeneity because of the sharp superconducting
transition width ($< 1$ K) revealed in the magnetization
measurement of our MgB$_2$ pellet. In other words, had the gap
variation been the result of the grain-to-grain stoichiometric
variation, we would have observed a very broad $T_c$ distribution
in the magnetization measurements, from $\sim 39$ K to $\sim 24$ K
for the $5 \sim 8$ meV gap variation. Given the quality of the
spectra and topography of our well characterized sample surfaces,
we suggest that \emph{the variation observed in the gap values of
$MgB_2$ pellets is the result of different grain orientations
relative to the incident quasiparticles}. The single gap value in
the c-axis oriented films further corroborates the notion of
$k$-dependent pairing potential. More importantly, had the pairing
symmetry been isotropic $s$-wave, the $(2 \Delta /k_B T_c)$ ratios
deduced from our tunneling spectra would not have varied from
$\sim 2.5$ to $\sim 4.5$ from grain to grain for $T_c$ variation
smaller than 1.0 K. In addition, to date there is no known theory
for isotropic $s$-wave superconductors that can justify a $(2
\Delta /k_B T_c)$ ratio smaller than the BCS value.

On the other hand, the electronic and structural anisotropy in the
MgB$_2$ system can lead to anisotropic $s$-wave pairing, and
therefore a $\vec k$-dependent pairing potential and a range of
gap values in the STS studies of polycrystalline samples.
Comparing the two possibilities of anisotropic s-wave pairing
potentials depicted in Figs.~\ref{fig1}(a) and \ref{fig1}(b), we
note that an in-plane 6-fold anisotropy would have resulted in a
c-axis spectrum with a sharp peak at $\Delta _0 (1 + \epsilon)$
and complicated spectral curvatures in $\mathcal{N} (E)$ for
$\Delta _0 (1 - \epsilon) < E < \Delta _0 (1 + \epsilon)$, as
shown in Fig.~\ref{fig1}(a). Such behavior was never seen in our
data. In contrast, spectra derived from the order parameter in
Fig.~\ref{fig1}(b) appeared to be most consistent with our finding
of smooth spectra on all samples, and with one maximum gap value
at $\Delta _z \sim 8$ meV for the c-axis films.

Using the anisotropic $s$-wave pairing potential $\Delta
_k~=~\Delta _{xy} \sin ^2 \theta _k + \Delta _z \cos ^2 \theta_k$,
with the minimum gap $\Delta _{xy} \approx 5$ meV and the maximum
gap $\Delta _z \approx 8$ meV determined empirically, we can
consistently account for all experimental data on both pellets and
c-axis films by varying one parameter $\theta _k$. As exemplified
in the main panel and inset of Fig.~\ref{fig3}(a), the former is
consistent with $\theta _k = (\pi/5)$ and the latter with $\theta
_k = 0$. Similarly, the same pairing potential can also be applied
to the c-axis film data with $\theta _k = 0$, as shown in the main
panel of Fig.~\ref{fig3}(b). Our empirical finding of a smaller
in-plane gap value ($\Delta _{xy} < \Delta _z$) is consistent with
the stronger in-plane Coulomb repulsion in MgB$_2$ \cite{ref24}. A
similar anisotropic $s$-wave pairing scenario has also been
proposed recently to account for the thermodynamic and optical
properties of MgB$_2$ wires \cite{ref25}. Furthermore, a number of
recent experimental reports, including the upper critical field
($H_{c2}$) measurements \cite{ref26,ref26b}, high-resolution
photoemission spectroscopy \cite{ref27}, and electron spin
resonance \cite{ref28}, are supportive of significantly
anisotropic properties in the superconducting state of MgB$_2$.

In summary, we have investigated the possible pairing channels in
MgB$_2$ based on group theory consideration, and have calculated
the quasiparticle spectra using a generalized BTK theory for
quasiparticle tunneling in 3D. Comparing the calculated results
with spectra taken on fully characterized MgB$_2$ pellets and
c-axis oriented films, we conclude that the order parameter of
MgB$_2$ belongs to the $A_{1g}$-representation of $D_{6h}$ group,
and is best described by an anisotropic $s$-wave pairing potential
with uniaxial symmetry.

\begin{acknowledgments}
The research at Caltech was supported by NSF Grant \#DMR-0103045 and
the Caltech President's Fund. Part of the work described in this
paper was performed by the Center for Space Microelectronics Technology,
Jet Propulsion Laboratory, and was sponsored by NASA. The work at
Pohang University was supported by the Ministry of Science and
Technology of Korea through the Creative Research Initiative Program.
\end{acknowledgments}


\begin{thebibliography}{32}
\expandafter\ifx\csname
natexlab\endcsname\relax\def\natexlab#1{#1}\fi
\expandafter\ifx\csname bibnamefont\endcsname\relax
  \def\bibnamefont#1{#1}\fi
\expandafter\ifx\csname bibfnamefont\endcsname\relax
  \def\bibfnamefont#1{#1}\fi
\expandafter\ifx\csname citenamefont\endcsname\relax
  \def\citenamefont#1{#1}\fi
\expandafter\ifx\csname url\endcsname\relax
  \def\url#1{\texttt{#1}}\fi
\expandafter\ifx\csname
urlprefix\endcsname\relax\def\urlprefix{URL }\fi
\providecommand{\bibinfo}[2]{#2}
\providecommand{\eprint}[2][]{\url{#2}}

\bibitem[{\citenamefont{Nagamatsu et~al.}(2001)\citenamefont{Nagamatsu,
  Nakagawa, Muranaka, Zenitani, and Akimitsu}}]{ref1}
\bibinfo{author}{\bibfnamefont{J.}~\bibnamefont{Nagamatsu}},
  \bibinfo{author}{\bibfnamefont{N.}~\bibnamefont{Nakagawa}},
  \bibinfo{author}{\bibfnamefont{T.}~\bibnamefont{Muranaka}},
  \bibinfo{author}{\bibfnamefont{Y.}~\bibnamefont{Zenitani}}, \bibnamefont{and}
  \bibinfo{author}{\bibfnamefont{J.}~\bibnamefont{Akimitsu}},
  \bibinfo{journal}{Nature} \textbf{\bibinfo{volume}{410}}, \bibinfo{pages}{63}
  (\bibinfo{year}{2001}).

\bibitem[{\citenamefont{Kotegawa et~al.}(2001)\citenamefont{Kotegawa, Ishida,
  Kitaoka, Muranaka, and Akimitsu}}]{ref2}
\bibinfo{author}{\bibfnamefont{H.}~\bibnamefont{Kotegawa}},
  \bibinfo{author}{\bibfnamefont{K.}~\bibnamefont{Ishida}},
  \bibinfo{author}{\bibfnamefont{Y.}~\bibnamefont{Kitaoka}},
  \bibinfo{author}{\bibfnamefont{T.}~\bibnamefont{Muranaka}}, \bibnamefont{and}
  \bibinfo{author}{\bibfnamefont{J.}~\bibnamefont{Akimitsu}}
  (\bibinfo{year}{2001}), \eprint{cond-mat/0102334}.

\bibitem[{\citenamefont{Bud'ko et~al.}(2001{\natexlab{a}})\citenamefont{Bud'ko,
  Lapertot, Petrovic, Cunningham, Anderson, and Canfield}}]{ref3}
\bibinfo{author}{\bibfnamefont{S.~L.} \bibnamefont{Bud'ko}},
  \bibinfo{author}{\bibfnamefont{G.}~\bibnamefont{Lapertot}},
  \bibinfo{author}{\bibfnamefont{C.}~\bibnamefont{Petrovic}},
  \bibinfo{author}{\bibfnamefont{C.~E.} \bibnamefont{Cunningham}},
  \bibinfo{author}{\bibfnamefont{N.}~\bibnamefont{Anderson}}, \bibnamefont{and}
  \bibinfo{author}{\bibfnamefont{P.~C.} \bibnamefont{Canfield}},
  \bibinfo{journal}{Phys. Rev. Lett.} \textbf{\bibinfo{volume}{86}},
  \bibinfo{pages}{1877} (\bibinfo{year}{2001}{\natexlab{a}}).

\bibitem[{\citenamefont{Rubio-Bollinger
  et~al.}(2001)\citenamefont{Rubio-Bollinger, Suderow, and Vieira}}]{ref4}
\bibinfo{author}{\bibfnamefont{G.}~\bibnamefont{Rubio-Bollinger}},
  \bibinfo{author}{\bibfnamefont{H.}~\bibnamefont{Suderow}}, \bibnamefont{and}
  \bibinfo{author}{\bibfnamefont{S.}~\bibnamefont{Vieira}},
  \bibinfo{journal}{Phys. Rev. Lett.} \textbf{\bibinfo{volume}{86}},
  \bibinfo{pages}{5582} (\bibinfo{year}{2001}).

\bibitem[{\citenamefont{Karapetrov et~al.}(2001)\citenamefont{Karapetrov,
  Iavarone, Kwok, Crabtree, and Hinks}}]{ref5}
\bibinfo{author}{\bibfnamefont{G.}~\bibnamefont{Karapetrov}},
  \bibinfo{author}{\bibfnamefont{M.}~\bibnamefont{Iavarone}},
  \bibinfo{author}{\bibfnamefont{W.~K.} \bibnamefont{Kwok}},
  \bibinfo{author}{\bibfnamefont{G.~W.} \bibnamefont{Crabtree}},
  \bibnamefont{and} \bibinfo{author}{\bibfnamefont{D.~G.} \bibnamefont{Hinks}},
  \bibinfo{journal}{Phys. Rev. Lett.} \textbf{\bibinfo{volume}{86}},
  \bibinfo{pages}{4374} (\bibinfo{year}{2001}).

\bibitem[{\citenamefont{Sharoni et~al.}(2001)\citenamefont{Sharoni, Felner, and
  Millo}}]{ref6}
\bibinfo{author}{\bibfnamefont{A.}~\bibnamefont{Sharoni}},
  \bibinfo{author}{\bibfnamefont{I.}~\bibnamefont{Felner}}, \bibnamefont{and}
  \bibinfo{author}{\bibfnamefont{O.}~\bibnamefont{Millo}},
  \bibinfo{journal}{Phys. Rev. B} \textbf{\bibinfo{volume}{63}},
  \bibinfo{pages}{220508} (\bibinfo{year}{2001}).

\bibitem[{\citenamefont{Schmidt et~al.}(2001)\citenamefont{Schmidt,
  Zasadzinski, Gray, and Hinks}}]{ref7}
\bibinfo{author}{\bibfnamefont{H.}~\bibnamefont{Schmidt}},
  \bibinfo{author}{\bibfnamefont{J.~F.} \bibnamefont{Zasadzinski}},
  \bibinfo{author}{\bibfnamefont{K.~E.} \bibnamefont{Gray}}, \bibnamefont{and}
  \bibinfo{author}{\bibfnamefont{D.~G.} \bibnamefont{Hinks}},
  \bibinfo{journal}{Phys. Rev. B} \textbf{\bibinfo{volume}{63}},
  \bibinfo{pages}{220504} (\bibinfo{year}{2001}).

\bibitem[{\citenamefont{Kortus et~al.}(2001)\citenamefont{Kortus, Mazin,
  Belashchenko, Antropov, and Boyer}}]{ref8}
\bibinfo{author}{\bibfnamefont{J.}~\bibnamefont{Kortus}},
  \bibinfo{author}{\bibfnamefont{I.~I.} \bibnamefont{Mazin}},
  \bibinfo{author}{\bibfnamefont{K.~D.} \bibnamefont{Belashchenko}},
  \bibinfo{author}{\bibfnamefont{V.~P.} \bibnamefont{Antropov}},
  \bibnamefont{and} \bibinfo{author}{\bibfnamefont{L.~L.} \bibnamefont{Boyer}},
  \bibinfo{journal}{Phys. Rev. Lett.} \textbf{\bibinfo{volume}{86}},
  \bibinfo{pages}{4656} (\bibinfo{year}{2001}).

\bibitem[{\citenamefont{An and Pickett}(2001)}]{ref9}
\bibinfo{author}{\bibfnamefont{J.~M.} \bibnamefont{An}} \bibnamefont{and}
  \bibinfo{author}{\bibfnamefont{W.~E.} \bibnamefont{Pickett}},
  \bibinfo{journal}{Phys. Rev. Lett.} \textbf{\bibinfo{volume}{86}},
  \bibinfo{pages}{4366} (\bibinfo{year}{2001}).

\bibitem[{\citenamefont{Kang et~al.}(2001{\natexlab{a}})\citenamefont{Kang,
  Jung, Kim, Park, Lee, Kim, Choi, Kim, Kim, and Lee}}]{ref10}
\bibinfo{author}{\bibfnamefont{W.~N.} \bibnamefont{Kang}},
  \bibinfo{author}{\bibfnamefont{C.~U.} \bibnamefont{Jung}},
  \bibinfo{author}{\bibfnamefont{K.~H.~P.} \bibnamefont{Kim}},
  \bibinfo{author}{\bibfnamefont{M.-S.} \bibnamefont{Park}},
  \bibinfo{author}{\bibfnamefont{S.~Y.} \bibnamefont{Lee}},
  \bibinfo{author}{\bibfnamefont{H.-J.} \bibnamefont{Kim}},
  \bibinfo{author}{\bibfnamefont{E.-M.} \bibnamefont{Choi}},
  \bibinfo{author}{\bibfnamefont{K.~H.} \bibnamefont{Kim}},
  \bibinfo{author}{\bibfnamefont{M.-S.} \bibnamefont{Kim}}, \bibnamefont{and}
  \bibinfo{author}{\bibfnamefont{S.-I.} \bibnamefont{Lee}}
  (\bibinfo{year}{2001}{\natexlab{a}}), \bibinfo{note}{submitted to Appl. Phys.
  Lett}, \eprint{cond-mat/0102313}.

\bibitem[{\citenamefont{Panagopoulos et~al.}(2001)\citenamefont{Panagopoulos,
  Rainford, Xiang, Scott, Kambara, and Inoue}}]{ref11}
\bibinfo{author}{\bibfnamefont{C.}~\bibnamefont{Panagopoulos}},
  \bibinfo{author}{\bibfnamefont{B.~D.} \bibnamefont{Rainford}},
  \bibinfo{author}{\bibfnamefont{T.}~\bibnamefont{Xiang}},
  \bibinfo{author}{\bibfnamefont{C.~A.} \bibnamefont{Scott}},
  \bibinfo{author}{\bibfnamefont{M.}~\bibnamefont{Kambara}}, \bibnamefont{and}
  \bibinfo{author}{\bibfnamefont{I.~H.} \bibnamefont{Inoue}}
  (\bibinfo{year}{2001}), \eprint{cond-mat/0103060}.

\bibitem[{\citenamefont{Vasquez et~al.}(2001)\citenamefont{Vasquez, Jung, Park,
  Kim, Kim, and Lee}}]{ref12}
\bibinfo{author}{\bibfnamefont{R.~P.} \bibnamefont{Vasquez}},
  \bibinfo{author}{\bibfnamefont{C.~U.} \bibnamefont{Jung}},
  \bibinfo{author}{\bibfnamefont{M.-S.} \bibnamefont{Park}},
  \bibinfo{author}{\bibfnamefont{H.-J.} \bibnamefont{Kim}},
  \bibinfo{author}{\bibfnamefont{J.~Y.} \bibnamefont{Kim}}, \bibnamefont{and}
  \bibinfo{author}{\bibfnamefont{S.-I.} \bibnamefont{Lee}},
  \bibinfo{journal}{Phys. Rev. B} \textbf{\bibinfo{volume}{64}},
  \bibinfo{pages}{052510} (\bibinfo{year}{2001}).

\bibitem[{\citenamefont{Yeh et~al.}(2001{\natexlab{a}})\citenamefont{Yeh, Wei,
  Chen, Si, and Xi}}]{ref13}
\bibinfo{author}{\bibfnamefont{N.-C.} \bibnamefont{Yeh}},
  \bibinfo{author}{\bibfnamefont{J.~Y.~T.} \bibnamefont{Wei}},
  \bibinfo{author}{\bibfnamefont{C.-T.} \bibnamefont{Chen}},
  \bibinfo{author}{\bibfnamefont{W.~D.} \bibnamefont{Si}}, \bibnamefont{and}
  \bibinfo{author}{\bibfnamefont{X.~X.} \bibnamefont{Xi}}
  (\bibinfo{year}{2001}{\natexlab{a}}).

\bibitem[{\citenamefont{Yeh et~al.}(2001{\natexlab{b}})\citenamefont{Yeh, Chen,
  Hammerl, Mannhart, Tajima, Yoshida, Schmehl, Schneider, and Schulz}}]{ref13b}
\bibinfo{author}{\bibfnamefont{N.-C.} \bibnamefont{Yeh}},
  \bibinfo{author}{\bibfnamefont{C.-T.} \bibnamefont{Chen}},
  \bibinfo{author}{\bibfnamefont{G.}~\bibnamefont{Hammerl}},
  \bibinfo{author}{\bibfnamefont{J.}~\bibnamefont{Mannhart}},
  \bibinfo{author}{\bibfnamefont{S.}~\bibnamefont{Tajima}},
  \bibinfo{author}{\bibfnamefont{K.}~\bibnamefont{Yoshida}},
  \bibinfo{author}{\bibfnamefont{A.}~\bibnamefont{Schmehl}},
  \bibinfo{author}{\bibfnamefont{C.~W.} \bibnamefont{Schneider}},
  \bibnamefont{and} \bibinfo{author}{\bibfnamefont{R.~R.} \bibnamefont{Schulz}}
  (\bibinfo{year}{2001}{\natexlab{b}}), \bibinfo{note}{to appear in Physica C
  (2001)}, \eprint{cond-mat/0103205}.

\bibitem[{\citenamefont{Wei et~al.}(1998)\citenamefont{Wei, Yeh, Garrigus, and
  Strasik}}]{ref14}
\bibinfo{author}{\bibfnamefont{J.~Y.~T.} \bibnamefont{Wei}},
  \bibinfo{author}{\bibfnamefont{N.-C.} \bibnamefont{Yeh}},
  \bibinfo{author}{\bibfnamefont{D.~F.} \bibnamefont{Garrigus}},
  \bibnamefont{and} \bibinfo{author}{\bibfnamefont{M.}~\bibnamefont{Strasik}},
  \bibinfo{journal}{Phys. Rev. Lett.} \textbf{\bibinfo{volume}{81}},
  \bibinfo{pages}{2542} (\bibinfo{year}{1998}).

\bibitem[{\citenamefont{Tsuei and Kirtley}(2000)}]{ref15}
\bibinfo{author}{\bibfnamefont{C.~C.} \bibnamefont{Tsuei}} \bibnamefont{and}
  \bibinfo{author}{\bibfnamefont{J.~R.} \bibnamefont{Kirtley}},
  \bibinfo{journal}{Rev. Mod. Phys.} \textbf{\bibinfo{volume}{72}},
  \bibinfo{pages}{969} (\bibinfo{year}{2000}), \bibinfo{note}{and references
  therein}.

\bibitem[{\citenamefont{VanHarlingen}(1995)}]{ref16}
\bibinfo{author}{\bibfnamefont{D.~L.} \bibnamefont{VanHarlingen}},
  \bibinfo{journal}{Rev. Mod. Phys.} \textbf{\bibinfo{volume}{67}},
  \bibinfo{pages}{515} (\bibinfo{year}{1995}), \bibinfo{note}{and references
  therein}.

\bibitem[{\citenamefont{Hu}(1994)}]{ref17}
\bibinfo{author}{\bibfnamefont{C.~R.} \bibnamefont{Hu}},
  \bibinfo{journal}{Phys. Rev. Lett.} \textbf{\bibinfo{volume}{72}},
  \bibinfo{pages}{1526} (\bibinfo{year}{1994}).

\bibitem[{\citenamefont{Tanaka and Kashiwaya}(1995)}]{ref18}
\bibinfo{author}{\bibfnamefont{Y.}~\bibnamefont{Tanaka}} \bibnamefont{and}
  \bibinfo{author}{\bibfnamefont{S.}~\bibnamefont{Kashiwaya}},
  \bibinfo{journal}{Phys. Rev. Lett.} \textbf{\bibinfo{volume}{74}},
  \bibinfo{pages}{3451} (\bibinfo{year}{1995}).

\bibitem[{\citenamefont{Blonder et~al.}(1982)\citenamefont{Blonder, Tinkham,
  and Klapwijk}}]{ref19}
\bibinfo{author}{\bibfnamefont{G.~E.} \bibnamefont{Blonder}},
  \bibinfo{author}{\bibfnamefont{M.}~\bibnamefont{Tinkham}}, \bibnamefont{and}
  \bibinfo{author}{\bibfnamefont{T.~M.} \bibnamefont{Klapwijk}},
  \bibinfo{journal}{Phys. Rev. B} \textbf{\bibinfo{volume}{25}},
  \bibinfo{pages}{4515} (\bibinfo{year}{1982}).

\bibitem[{\citenamefont{Sung et~al.}(2001)\citenamefont{Sung, Kim, Kim, Yoo,
  Lee, Lee, Jung, Park, Kang, Zhonglian et~al.}}]{ref20}
\bibinfo{author}{\bibfnamefont{G.~Y.} \bibnamefont{Sung}},
  \bibinfo{author}{\bibfnamefont{S.~H.} \bibnamefont{Kim}},
  \bibinfo{author}{\bibfnamefont{J.~H.} \bibnamefont{Kim}},
  \bibinfo{author}{\bibfnamefont{D.~C.} \bibnamefont{Yoo}},
  \bibinfo{author}{\bibfnamefont{J.~W.} \bibnamefont{Lee}},
  \bibinfo{author}{\bibfnamefont{J.~Y.} \bibnamefont{Lee}},
  \bibinfo{author}{\bibfnamefont{C.~U.} \bibnamefont{Jung}},
  \bibinfo{author}{\bibfnamefont{M.-S.} \bibnamefont{Park}},
  \bibinfo{author}{\bibfnamefont{W.~N.} \bibnamefont{Kang}},
  \bibinfo{author}{\bibfnamefont{D.}~\bibnamefont{Zhonglian}},
  \bibnamefont{et~al.} (\bibinfo{year}{2001}), \eprint{cond-mat/0102498}.

\bibitem[{\citenamefont{Jung et~al.}(2001{\natexlab{a}})\citenamefont{Jung,
  Park, Kang, Kim, Kim, Lee, and Lee}}]{ref21}
\bibinfo{author}{\bibfnamefont{C.~U.} \bibnamefont{Jung}},
  \bibinfo{author}{\bibfnamefont{M.-S.} \bibnamefont{Park}},
  \bibinfo{author}{\bibfnamefont{W.~N.} \bibnamefont{Kang}},
  \bibinfo{author}{\bibfnamefont{M.-S.} \bibnamefont{Kim}},
  \bibinfo{author}{\bibfnamefont{K.~H.~P.} \bibnamefont{Kim}},
  \bibinfo{author}{\bibfnamefont{S.~Y.} \bibnamefont{Lee}}, \bibnamefont{and}
  \bibinfo{author}{\bibfnamefont{S.-I.} \bibnamefont{Lee}},
  \bibinfo{journal}{Appl. Phys. Lett.} \textbf{\bibinfo{volume}{78}},
  \bibinfo{pages}{4157} (\bibinfo{year}{2001}{\natexlab{a}}).

\bibitem[{\citenamefont{Jung et~al.}(2001{\natexlab{b}})\citenamefont{Jung,
  Park, Kang, Kim, Lee, and Lee}}]{ref21b}
\bibinfo{author}{\bibfnamefont{C.~U.} \bibnamefont{Jung}},
  \bibinfo{author}{\bibfnamefont{M.-S.} \bibnamefont{Park}},
  \bibinfo{author}{\bibfnamefont{W.~N.} \bibnamefont{Kang}},
  \bibinfo{author}{\bibfnamefont{M.-S.} \bibnamefont{Kim}},
  \bibinfo{author}{\bibfnamefont{S.~Y.} \bibnamefont{Lee}}, \bibnamefont{and}
  \bibinfo{author}{\bibfnamefont{S.-I.} \bibnamefont{Lee}},
  \bibinfo{journal}{Physica C} \textbf{\bibinfo{volume}{353}},
  \bibinfo{pages}{162} (\bibinfo{year}{2001}{\natexlab{b}}).

\bibitem[{\citenamefont{Kang et~al.}(2001{\natexlab{b}})\citenamefont{Kang,
  Kim, Choi, Jung, and Lee}}]{ref22}
\bibinfo{author}{\bibfnamefont{W.~N.} \bibnamefont{Kang}},
  \bibinfo{author}{\bibfnamefont{H.-J.} \bibnamefont{Kim}},
  \bibinfo{author}{\bibfnamefont{E.-M.} \bibnamefont{Choi}},
  \bibinfo{author}{\bibfnamefont{C.~U.} \bibnamefont{Jung}}, \bibnamefont{and}
  \bibinfo{author}{\bibfnamefont{S.-I.} \bibnamefont{Lee}},
  \bibinfo{journal}{Science} \textbf{\bibinfo{volume}{292}},
  \bibinfo{pages}{1521} (\bibinfo{year}{2001}{\natexlab{b}}).

\bibitem[{\citenamefont{Giubileo et~al.}(2001)\citenamefont{Giubileo,
  Roditchev, Sacks, Lamy, and Klein}}]{ref22b}
\bibinfo{author}{\bibfnamefont{F.}~\bibnamefont{Giubileo}},
  \bibinfo{author}{\bibfnamefont{D.}~\bibnamefont{Roditchev}},
  \bibinfo{author}{\bibfnamefont{W.}~\bibnamefont{Sacks}},
  \bibinfo{author}{\bibfnamefont{R.}~\bibnamefont{Lamy}}, \bibnamefont{and}
  \bibinfo{author}{\bibfnamefont{J.}~\bibnamefont{Klein}}
  (\bibinfo{year}{2001}), \eprint{cond-mat/0105146}.

\bibitem[{\citenamefont{Dynes et~al.}(1978)\citenamefont{Dynes, Narayanamurti,
  and Garno}}]{ref23}
\bibinfo{author}{\bibfnamefont{R.~C.} \bibnamefont{Dynes}},
  \bibinfo{author}{\bibfnamefont{V.}~\bibnamefont{Narayanamurti}},
  \bibnamefont{and} \bibinfo{author}{\bibfnamefont{J.~P.} \bibnamefont{Garno}},
  \bibinfo{journal}{Phys. Rev. Lett.} \textbf{\bibinfo{volume}{41}},
  \bibinfo{pages}{1509} (\bibinfo{year}{1978}).

\bibitem[{\citenamefont{Voelker et~al.}(2001)\citenamefont{Voelker, Anisimov,
  and Rice}}]{ref24}
\bibinfo{author}{\bibfnamefont{K.}~\bibnamefont{Voelker}},
  \bibinfo{author}{\bibfnamefont{V.~I.} \bibnamefont{Anisimov}},
  \bibnamefont{and} \bibinfo{author}{\bibfnamefont{T.~M.} \bibnamefont{Rice}}
  (\bibinfo{year}{2001}), \eprint{cond-mat/0103082}.

\bibitem[{\citenamefont{Haas and Maki}(2001)}]{ref25}
\bibinfo{author}{\bibfnamefont{S.}~\bibnamefont{Haas}} \bibnamefont{and}
  \bibinfo{author}{\bibfnamefont{K.}~\bibnamefont{Maki}}
  (\bibinfo{year}{2001}), \eprint{cond-mat/0104207}.

\bibitem[{\citenamefont{Xu et~al.}(2001)\citenamefont{Xu, Kitazawa, Takano, Ye,
  Nishida, Abe, Matsushita, and Kido}}]{ref26}
\bibinfo{author}{\bibfnamefont{M.}~\bibnamefont{Xu}},
  \bibinfo{author}{\bibfnamefont{H.}~\bibnamefont{Kitazawa}},
  \bibinfo{author}{\bibfnamefont{Y.}~\bibnamefont{Takano}},
  \bibinfo{author}{\bibfnamefont{J.}~\bibnamefont{Ye}},
  \bibinfo{author}{\bibfnamefont{K.}~\bibnamefont{Nishida}},
  \bibinfo{author}{\bibfnamefont{H.}~\bibnamefont{Abe}},
  \bibinfo{author}{\bibfnamefont{A.}~\bibnamefont{Matsushita}},
  \bibnamefont{and} \bibinfo{author}{\bibfnamefont{G.}~\bibnamefont{Kido}}
  (\bibinfo{year}{2001}), \eprint{cond-mat/0105271}.

\bibitem[{\citenamefont{Bud'ko et~al.}(2001{\natexlab{b}})\citenamefont{Bud'ko,
  Kogan, and Canfield}}]{ref26b}
\bibinfo{author}{\bibfnamefont{S.~L.} \bibnamefont{Bud'ko}},
  \bibinfo{author}{\bibfnamefont{V.~G.} \bibnamefont{Kogan}}, \bibnamefont{and}
  \bibinfo{author}{\bibfnamefont{P.~C.} \bibnamefont{Canfield}}
  (\bibinfo{year}{2001}{\natexlab{b}}), \eprint{cond-mat/0106577}.

\bibitem[{\citenamefont{Tsuda et~al.}(2001)\citenamefont{Tsuda, Yokoya, Kiss,
  Y.Takano, Togano, Kitou, Ihara, and Shin}}]{ref27}
\bibinfo{author}{\bibfnamefont{S.}~\bibnamefont{Tsuda}},
  \bibinfo{author}{\bibfnamefont{T.}~\bibnamefont{Yokoya}},
  \bibinfo{author}{\bibfnamefont{T.}~\bibnamefont{Kiss}},
  \bibinfo{author}{\bibnamefont{Y.Takano}},
  \bibinfo{author}{\bibfnamefont{K.}~\bibnamefont{Togano}},
  \bibinfo{author}{\bibfnamefont{H.}~\bibnamefont{Kitou}},
  \bibinfo{author}{\bibfnamefont{H.}~\bibnamefont{Ihara}}, \bibnamefont{and}
  \bibinfo{author}{\bibfnamefont{S.}~\bibnamefont{Shin}}
  (\bibinfo{year}{2001}), \eprint{cond-mat/0104489}.

\bibitem[{\citenamefont{Simon et~al.}(2001)\citenamefont{Simon, Janossy, Feher,
  Muranyi, Garaj, Forro, Petrovic, Bud'ko, Lapertot, Kogan et~al.}}]{ref28}
\bibinfo{author}{\bibfnamefont{F.}~\bibnamefont{Simon}},
  \bibinfo{author}{\bibfnamefont{A.}~\bibnamefont{Janossy}},
  \bibinfo{author}{\bibfnamefont{T.}~\bibnamefont{Feher}},
  \bibinfo{author}{\bibfnamefont{F.}~\bibnamefont{Muranyi}},
  \bibinfo{author}{\bibfnamefont{S.}~\bibnamefont{Garaj}},
  \bibinfo{author}{\bibfnamefont{L.}~\bibnamefont{Forro}},
  \bibinfo{author}{\bibfnamefont{C.}~\bibnamefont{Petrovic}},
  \bibinfo{author}{\bibfnamefont{S.~L.} \bibnamefont{Bud'ko}},
  \bibinfo{author}{\bibfnamefont{G.}~\bibnamefont{Lapertot}},
  \bibinfo{author}{\bibfnamefont{V.~G.} \bibnamefont{Kogan}},
  \bibnamefont{et~al.}, \bibinfo{journal}{Phys. Rev. Lett.}
  \textbf{\bibinfo{volume}{87}}, \bibinfo{pages}{047002}
  (\bibinfo{year}{2001}).

\end{thebibliography}
\end{document}